\documentclass[twocolumn]{aastex62}

\usepackage{apjfonts}
\usepackage[T1]{fontenc}
\usepackage{color}

\usepackage{amsmath,amstext}
\usepackage{hyperref}
\usepackage{natbib}
\newcommand\beq{\begin{equation}}
\newcommand\eeq{\end{equation}}

\shortauthors{Perna, Lazzati \& Farr}

\begin{document}

\title{LIMITS ON ELECTROMAGNETIC COUNTERPARTS OF GRAVITATIONAL WAVE-DETECTED BINARY BLACK HOLE MERGERS}

\author{Rosalba Perna}
\affiliation{Department of Physics and Astronomy, Stony Brook University, Stony Brook, NY, 11794, USA}
\affiliation{Center for Computational Astrophysics, Flatiron Institute, 162 5th Avenue, New York, NY 10010, \
USA}

\author{Davide Lazzati}
\affiliation{Department of Physics, Oregon State University, 301
Weniger Hall, Corvallis, OR 97331, USA}

\author{Will Farr}
\affiliation{Department of Physics and Astronomy, Stony Brook University, Stony Brook, NY, 11794, USA}
\affiliation{Center for Computational Astrophysics, Flatiron Institute, 162 5th Avenue, New York, NY 10010, \
USA}

\begin{abstract}
Unlike mergers of two compact objects containing a neutron star (NS),
binary black hole (BBH) mergers	are not	accompanied by the production
of tidally disrupted material, and hence lack the most direct source
of accretion	to power a jet and generate electromagnetic (EM)
radiation. However, following a	tentative detection by the Fermi GBM
of a $\gamma$-ray counterpart to GW150914, several  ideas
were proposed for driving a jet and producing EM radiation. {\em If}
such jets were in fact produced, they would however lack	the  cocoon emission
that makes jets	from binary NSs	bright also at large viewing angles.
Here, via Monte	Carlo simulations of a population of BBH mergers
with properties	consistent with those inferred from the existing LIGO/Virgo observations, and the angular
emission characteristic of jets propagating	into the interstellar medium,
we derive limits on the allowed energetics and Lorentz factors of such jets	from
EM follow ups to GW-detected BBH merger events to date, and we make predictions which will help tighten
these limits with broadband EM follow ups to  events in future LIGO/Virgo runs.
The condition that $\lesssim 1$ event out of 10 GW-detected BBH mergers be above the Fermi/GBM threshold imposes that any currently allowed emission model has to satisfy the condition $(E_{\rm iso}/10^{49}{\rm erg})(\theta_{\rm jet}/20^\circ)\lesssim 1$.

\end{abstract}

\keywords{gravitational waves ---  black hole physics --- gamma rays:bursts}

\section{Introduction}
The discovery of gravitational waves \citep{Abbott2016a} has opened a new window onto the Universe. Furthermore, the simultaneous detection of electromagnetic (EM) radiation from the double binary neutron star merger GW170817 \citep{Abbott2017c} has demonstrated the impact of these observations in several disparate areas of physics and astrophysics, from high energy astrophysics, to nuclear physics, to cosmology.

The general thinking is that EM radiation accompanying GWs from binary compact object mergers
requires at least one of the two objects to be a NS, whose tidally disrupted material
 provides the accretion energy required to power an electromagnetic counterpart.
However, following the first GW detection from a binary BH merger, the GBM detector on the Fermi satellite detected a tentative $\gamma$-ray counterpart, within 1~sec after the GW detection \citep{Connaughton2016}. While this was a low-statistics event, it was of enough interest to spur ideas that could explain such emission, if indeed real.
Since there is no accretion material resulting from tidal disruption at merger, various astrophysical scenarios were put forward in which the binary black holes (BBHs) would have some source of pre-existing material, whether related to the progenitor star \citep{Loeb2016,Woosley2016,Janiuk2017} and the mini-disk resulting from its supernova explosion
\citep{Perna2016,Murase2016,DeMink2017,Martin2018}, or to the environment of the merger, such as an AGN disk \citep{Bartos2017}.
Alternatively, the energy source could be entirely of eletromagnetic nature if the BHs are charged \citep{Liebling2016,Zhang2016,Liu2016,Fraschetti2018}.
GRMHD simulations have additionally demonstrated that
  jets are produced from merging BHs if there is some matter around the BHs at the time of merger \citep{Khan2018}.

The mere possibility that any of the scenarios above (or others) could be realized in nature is very interesting and worth testing.
As more GWs from BBH mergers are detected, and EM followups are conducted, 
the question is what constraints they
put to EM emission  models. 
The answer to this question depends on the physical characteristics of the emission (and in particular its geometrical beaming, total energy, Lorentz factor), on the distribution of properties of detected GW events as a function of redshift, and on the observer viewing angle with respect to the emitting jet.
An important difference with the case of a double NS merger (or NS-BH for small mass ratios) is that, even if both events were to produce a jet, 
in the former case the jet would be interacting with the ejecta from the tidally disrupted NS and produce the so-called 'cocoon' emission, bright at relatively wide angles
as observed in GW170817 \citep{Abbott2017c}. On the other hand,
 in the case of BBH mergers there are
no ejecta for any hypothetical jet to interact with, and hence the probability of observing EM radiation off axis is much lower, and dependent on both energy and Lorentz factor of the jet, in addition to the jet size.

In this paper we simulate
the evolution of jets expanding into a pure interstellar medium, without any interaction with ejecta from tidally disrupted material. We compute the time-dependent angular emission in 
$\gamma$-rays (prompt emission) at early times, as well as the longer-wavelength radiation (afterglow) naturally produced by dissipation of the jet into the medium (details in \S2). 
Given the distribution of redshifts and orbital inclinations of BBH mergers as deduced from the GW events observed to date (described in \S3), we perform Monte Carlo simulations to predict the fraction of events which would be expected to be above the threshold flux of currently observing instruments in typical observation bands, for a range of jet properties (energy, Lorentz factor, opening angle) (\S4). 
We further derive limits on the presence of jets and their properties (\S5) using the available data from the first LIGO run and the EM follow-ups to GW detections. We summarize and conclude in \S6.

\section{Prompt and afterglow emission from a jet without cocoon}

The wide-angle emission, both in the prompt phase and during the
afterglow, is computed using the formalism of \citet{Lazzati2017a}\footnote{This is a simplified formalism compared to the full hydrodynamical jet simulations of \citet{Lazzati2017b} and \citet{Lazzati2018}. However, it allows us to explore a wider range of jet parameters due to shorter simulation times, while preserving the main features of the jet evolution.}.
Two counter-propagating jets, of energy $E/2$ each, initial Lorentz
factor $\Gamma_0$, and uniform properties within an angle $\theta_j$,
propagate into an interstellar medium of density $n$. 
The prompt emission is computed assuming that the internal energy of
the outflow is dissipated at some distance $R_{\rm rad}$ from the engine,
and that the duration of the emission lasts for a certain time $t^\prime_{\rm eng}$ in the comoving frame of the outflow.
The observed bolometric flux is obtained by integrating the local emission
over the entire emitting surface, after boosting by the forth power of the Doppler factor 
$\delta(\Gamma,\theta)=\left[\left(1\,-\,\beta\cos\theta\right)\right]^{-1}$, where
$\beta$ is the jet speed in units of the speed of light, and $\theta$ is the angle
that the outgoing photon makes with the normal to the jet surface.

For the emission in a specific energy band, we assume a Band spectrum
\citep{Band1997} with { spectral} power-law indices $\alpha_{\rm
  ph}=0$ and $\beta_{\rm ph}=-2.5$, and comoving peak frequency
$h\nu^\prime_{\rm pk}=2.5$~keV. {These gives a typical GRB spectrum
  \citep{Gruber2014} with observed peak frequency of 500 keV (for
  $\Gamma=100$) or an X-ray flash with peak frequency of 50 keV (for
  $\Gamma=10$).} Light curves are calculated by adding up the
radiation from three million emission regions, each activated at its
own $R_{\rm rad}$, and reaching the observer at a time depending on
both the production radius, as well as as the time delay in reaching
the observer.

\begin{figure}
\includegraphics[width=9cm]{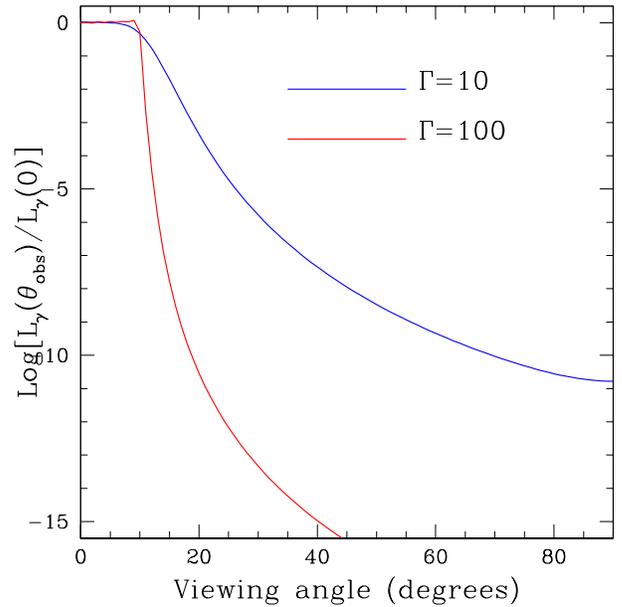}
\vspace{-0.6in}
\caption{Peak luminosity in the Fermi/GBM band as a function of the observer viewing angle $\theta_{\rm obs}$  with respect to the axis of the closest jet. The function is normalized to its value at $\theta_{\rm obs}=0$, i.e. when the jet is observed on-axis.}
\label{fig:prompt}
\end{figure}

For the  typical model that we study in this paper (but see later for extensions), we adopt a fiducial value of the jet opening
angle of $\theta_{\rm jet}=10^\circ$, and explore six different
models, produced by the combination of three energy values for the jet, $E=10^{46}, 10^{47}, 10^{48}$~ergs and two initial values for the
Lorentz factor of the jet, $\Gamma=10,100$.  Note that these are the actual energies, which, for the chosen jet angle of $10^\circ$, correspond to isotropic energies $\sim 65$ times higher. These isotropic equivalent values straddle the energy inferred for the Fermi/GBM candidate counterpart ($E_{\rm iso}\sim 10^{49}$~ergs). For the Lorentz factor, on the other hand, the two chosen values represent a highly relativistic jet and a mildly relativistic one.  Due to relativistic Doppler beaming, the variation of the brightness with viewing angle is very sensitive to the Lorentz factor of the jet, as shown in Fig.~\ref{fig:prompt}.

The afterglow radiation, from the X-ray to the radio band, is
produced as the jet drives a relativistic shock that propagates and dissipates into the interstellar
medium. The local emission is synchrotron radiation \citep{Sari1998},
and the total observed spectrum is computed using a semi-analytic afterglow code (see, e.g., Rossi et al. 2004; Lazzati et al. 2018). The code
describes the emission of a relativistic fireball with an arbitrary
energy distribution, as seen by observers at arbitrary viewing angles
with respect to the jet axis.

\begin{figure*}
\includegraphics[width=9cm]{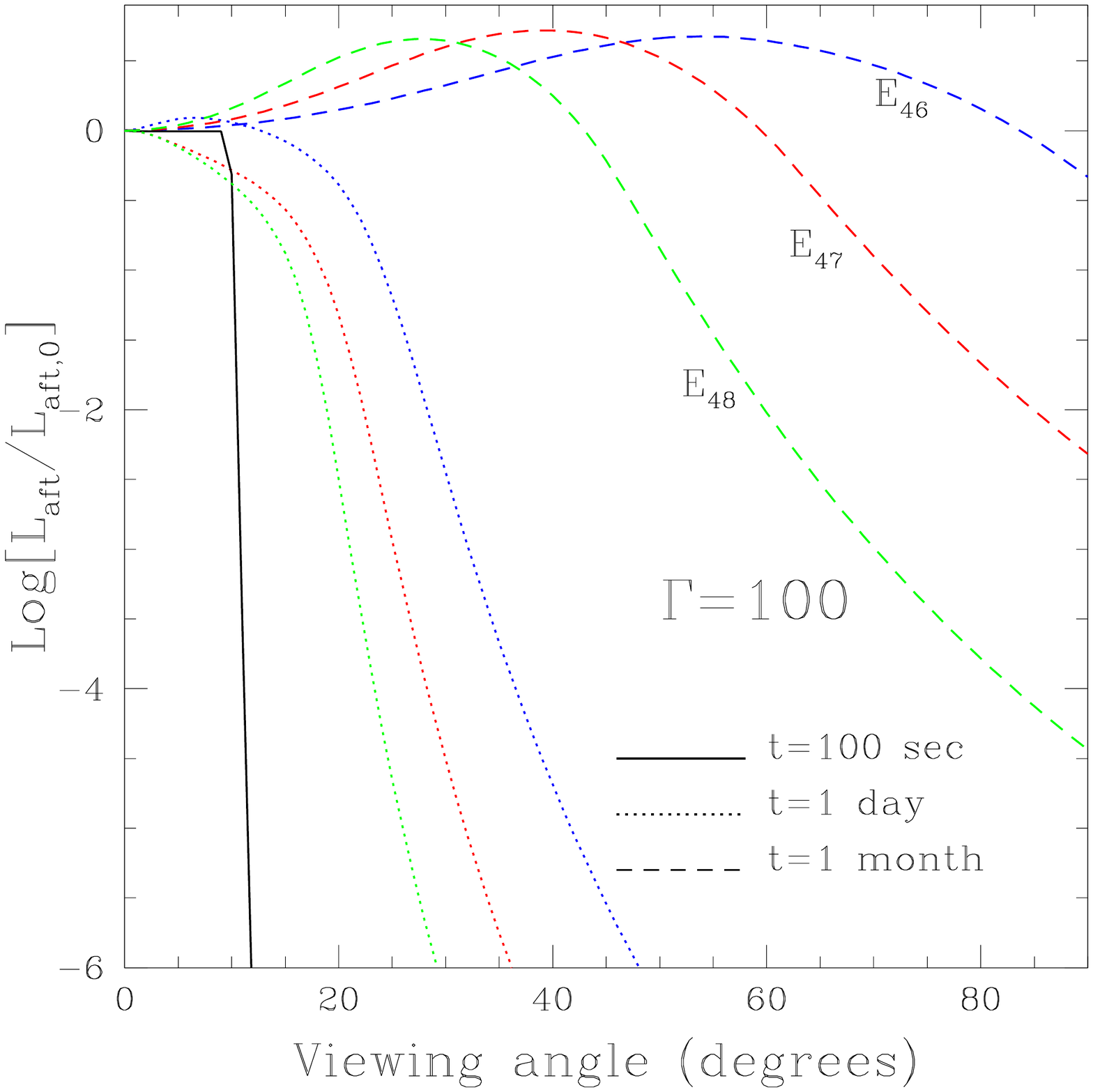}
\includegraphics[width=9cm]{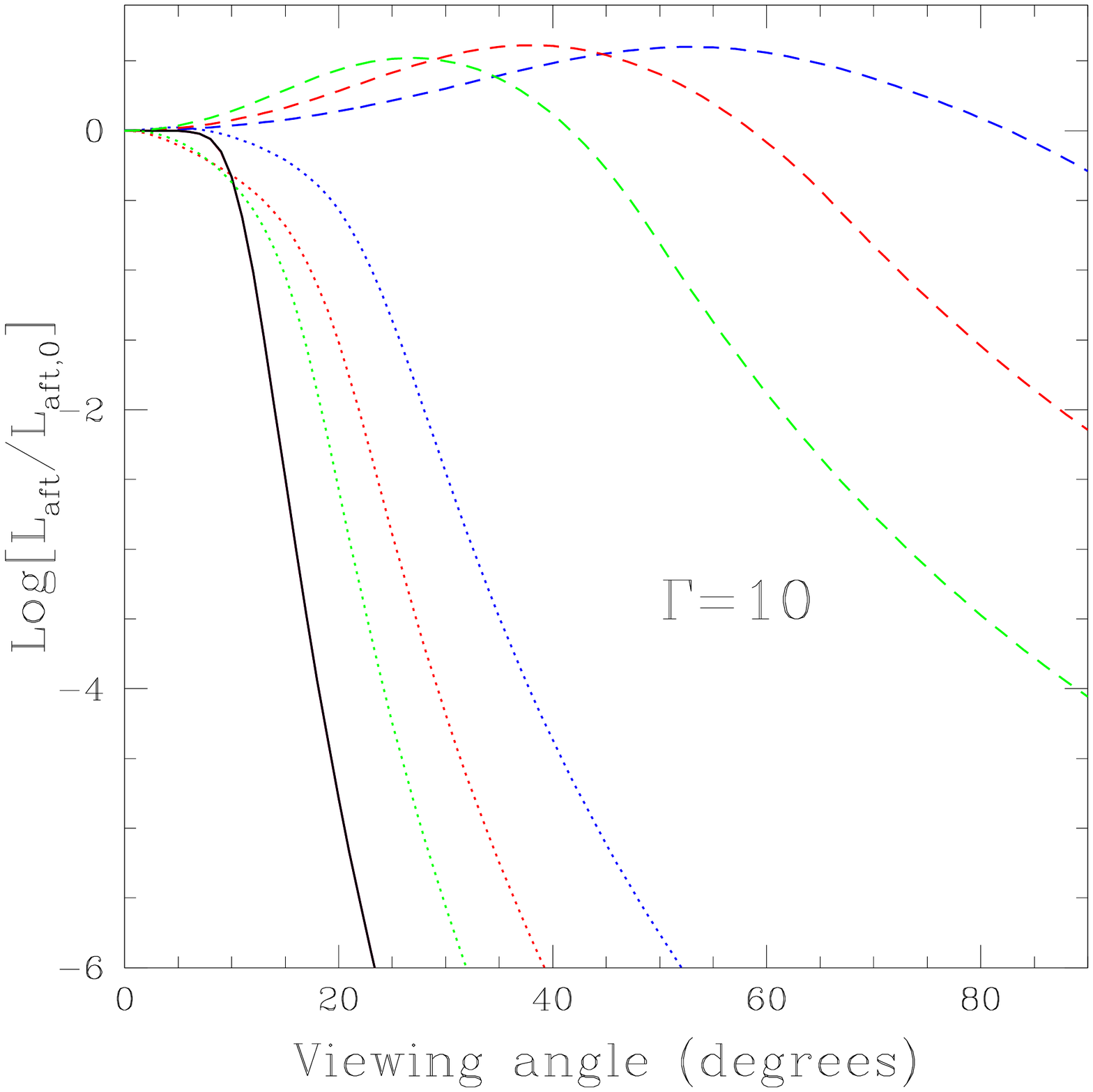}
\vspace{-0.85in}
\caption{Afterglow luminosity as a function of the observer viewing angle
$\theta_{\rm obs}$ with respect to the axis of the closest jet, for three times
and the three energy values studied here. In each case,
the luminosity function is normalized to its value at $\theta_{\rm obs}=0$, i.e. when the jet is observed on-axis. Note that,
while the colors distinguish the three energy values ($E_{46}$ in blue, $E_{47}$ in red, and $E_{48}$ in green), at the two later times (1 day and 1 month), at $t=100$~sec, the (normalized) curves are the same independently of energy, and they are displayed with a single black line for clarity. More energetic jets take a longer time to slow down and hence to isotropize.  }
\vspace{0.3in}
\label{fig:aft_teta}
\end{figure*}

In describing the results of our event simulation in \S4, we will
indicate with the corresponding subscripts the three energy values and
the two Lorentz factors, so that, for example, model
$E_{47}\Gamma_{10}$ would correspond to the energy $E_{\rm
  iso}=10^{47}$~ergs, and to the Lorentz factor $\Gamma=10$. Note that
the prompt emission scales linearly with energy, and so does the peak
of the afterglow emission; however, the afterglow radiation at some
specific frequency (or in a given band), does generally not since the
break frequencies where the spectrum changes shape depend on energy
\citep{Sari1998}.  The afterglow intensity further depends on the
medium ambient density $n$, as well as on the fraction $\epsilon_e$
and $\epsilon_B$ of jet energy that goes into the electrons and the
magnetic field, respectively, {and on the number fraction of
  accelerated electrons.  Inferences of the values of these parameters
  have been derived via dedicated broadband modeling of the afterglows
  of gamma-ray bursts. For the astrophysical scenario that we are
  studying here, the most appropriate sample for comparison is the
  comprehensive catalogue of 103 short GRBs with prompt follow-up
  observations in the X-ray, optical, near-infrared and/or radio
  bands.  Fits to the lightcurve of each burst were performed by
  \citet{Fong2015}.  Due to the lack of a continuous monitoring in
  most events, they could not simultaneously fit for all the model
  parameters, and hence kept fixed the parameter $\epsilon_e=0.1$. The
  magnetic field energy fraction $\epsilon_B$ was then found to be
  consistent with either 0.01 or 0.1 for the greatest majority of the
  bursts (only a few outliers required smaller values).  Since the
  basic physics of the afterglow phenomenology is expected to be
  similar for long and short GRBs (it is the result of a point
  explosion in an external medium), it is useful to look also at
  results of broadband modeling for long GRBs, which, being typically
  brighter, generally have a more complete set of broadband data. The
  sample of well-monitored GRBs modeled by \citet{Panaitescu2001} was
  found to have $\epsilon_B$ varying from a few $\times 10^{-5}$ to a
  few $\times 10^{-1}$, with the largest number of bursts concentrated
  in the higher range (and those higher values have smaller error
  bars). The value of $\epsilon_e$ was found to be clustered between
  $\sim 0.01-0.1$.  On the other hands, analysis of other bursts by
  different groups have found lower values; i.e. \citet{Wang2015}
  found, for a fixed $\epsilon_e=0.1$ in their fits, that their sample
  had $\epsilon_B\lesssim 10^{-3}$.

From a theoretical point of view, if the shock simply compresses the upstream magnetic field, then $\epsilon_B$ is expected to be low, on the order of $\sim 10^{-7}-10^{-6}$. On the other hand, if the magnetic field is amplified at the shock front via plasma instabilities, then  $\epsilon_B$ can be as high as $\sim 0.1$ \citep{Medvedev1999,Nishikawa2009}. 
In our simulations, we adopt}
 $\epsilon_e=0.03$ and $\epsilon_B=0.01$. {To zeroth order, afterglow luminosities for different values of these parameters can then be derived via analytical scalings \citep{Sari1998}. 
 Following customary habits in afterglow modeling, we further assume that the fraction of electrons which undergo acceleration is on the order of 1. This parameter is hardly constrained by observations, and it is highly degenerate with the other microphysical parameters of the shock (see i.e. discussion in \citealt{Eichler2005}).  
 }

The number density, on the other hand, will depend on the type of
galaxy and size in which the merger events occur.  {This (external)
  variable is expected to vary with the type of progenitors, since
  merger locations depend on the progenitor type. For BBH mergers,
  location sites are completely unknown from an observational point of
  view, at least to date. However, there have been recent numerical
  (i.e. population synthesis) simulations of isolated binary evolution
  tracking merger locations \citep{Perna2018} that can provide a
  guide. These have shown that, for large galaxies, most of the events
  will occur in environments with number densities between $\sim
  10^{-4}-1$~cm$^{-3}$, with generally larger values expected for
  spiral galaxies and lower for elliptical ones.  In our simulations
  we assume $n= 0.01$~cm$^{-1}$ as a mean, representative
  value. However, likewise for the other parameters described above,
  the luminosities that we derive can be roughly rescaled analytically
  to different density values, if needed.

The jet	size, on the other hand, strongly influences the magnitude of
the angular luminosity, which is of fundamental importance for the
study carried out here.  General relativistic hydrodynamic simulations
of accretion flows around black holes remnants of compact object
mergers \citep{Aloy2005} show that ultrarelativistic jets can be
driven by thermal energy deposition (possibly due to
neutrino-antineutrino annihilation), for energy	deposition rates above
about $10^{48}$~erg~s$^{-1}$ and sufficientlhy low baryon density. In
those simulations,
jets are found to have opening angles $\sim 5^\circ-10^\circ$, and a sharp
edge embedded laterally	by a wind with a steeply declining Lorentz factor.
Alternatively, jets could be powered via the Blandford-Znajek (BZ) mechanism \citep{Blandford1977}. General relativistic magnetohydrodynamic simulations \citep{Kathirg2019} of jets propagating in the environment expected post-merger from a binary neutron star system find a roughly constant Lorentz factor of $\sim 100$ within an angle of about $10^\circ$, dropping very rapidly at larger angles. Almost all of the energy is concentrated within $< 10^\circ$. The luminosity, while also dropping steeply (and becoming $\lesssim 10^{-4}$ of the maximum at viewing angles larger than $\sim 30^\circ$), is however shallower than it would have been for a top-hat jet, due to the interaction with the disk-wind. These simulations were however tailored to explain the properties of GRB170817A, resulting from a double neutron star merger, and the jet properties depend on the assumed properties of the disk-wind. For the binary BH case of interest here,  \citet{Khan2018} performed general relativistic magneto-hydrodynamic simulations of disk accretion onto black holes with a mass ratio similar to that measured for GW150914. They explored different disk models (in size, scale height), and found that collimated and magnetically dominated outflows emerge in the disk funnel independently of the properties of the disk. The Poynting luminosity is found to converge to the BZ value once quasi-equilibrium is reached. For a fiducial value $\eta=0.1$ of accretion efficiency, they found that an isotropic energy of 1.8$\times 10^{49}$~erg (as inferred for the candidate $\gamma$-ray counterpart to GW150914) can be achieved for a range of disk masses $\sim 10^{-4}-10^{-3}M_\odot$, with the specific value depending on the disk model. A fossil disk with mass
$\sim 10^{-4}M_\odot$ was discussed as a possibility for a dead disk formed from fallback after a supernova explosion \citep{Perna2016}. Hence, at least in theory, the conditions for generating jets from binary black hole mergers do exist. 
On the other hand, the specific angular dependence of the brightness of such jets will depend on their main driving mechanism, as well as on the structure of the associated disk. 
Given the above model dependencies, and the fact that bright lateral emission due to the interaction of the jet with ejecta
(producing the so-called  cocoon and/or a structured jet)
is not expected in the binary black hole merger scenario (though there can be weaker off-axis emission due to interaction with a disk-wind), here we adopt the simplest assumption of a top-hat jet with sharp edges. Should any additional emission be present due to disk-wind interaction, this would mostly affect the very large angles in models with high $\Gamma$ and at especially at early times, when $1/\Gamma$ is still small. Hence our results should be intended as {\em the most conservative ones} (for the given microphysical  parameters) in terms of observability. 
They will also be the most direct ones to use by observers, when only upper limits to the emission are available.}

Fig.~\ref{fig:aft_teta} shows the afterglow luminosity as a function of the
viewing angle at three representative times in the source frame:
$t=100$~ sec, 1 day, 1 month.  The difference between the $\Gamma=100$ (left panel) and
the $\Gamma=10$ (right panel) cases is especially apparent at early
times, when the fireball has not slowed down yet, and hence the
Doppler beaming of the radiation within $1/\Gamma$ causes a sharper
decline at viewing angles $\theta_{\rm obs}\gtrsim\theta_{\rm jet}$ for
the higher $\Gamma$ case.  The behaviour at later times, as a function
of the various model parameters, can be readily understood by
considering that the fireball starts to decelerate when it has
collected an amount of interstellar mass $M_{\rm ISM}\approx
E/\Gamma^2/c^2$. Therefore, for the same energy, the faster fireball
will slow down  at earlier times than the
initially-slower counterpart. On the other hand, for the same initial
$\Gamma$, the less energetic is the initial shock, the earlier it
slows down and its emission becomes isotropic with viewing angle.  Note
that, once $\Gamma\sim 1$, the brightness is not uniform with
$\theta_{\rm jet}$, as one may naively expect. This is due to the fact
that, unlike the prompt emission, which is produced for a thin shell
right where the fireball becomes optically thin, the afterglow comes
from a large radial region of optically thin material.  The radiation
that the observer receives depends on both the emission time at
location $R_{\rm em}$, as well as on the time for those photons to travel
to the observer.  For off-axis observers, there is an additional time
delay due to the additional path length of the radiation from the edge
of the jet closest to the observer, $t_{\rm del}=R_{\rm em}[1-\cos(\theta_{\rm
obs}-\theta_{\rm jet})]$, which is longer at larger $\theta_{\rm
obs}$.  Therefore, emission at larger viewing angles has a
contribution from earlier times in the frame of the fireball, when the
fireball was brighter.  This would explain why there are ranges of viewing geometries for which observers at larger
angular distances from the jet axis can measure a brighter emission than observers at smaller angles.
For the same initial $\Gamma$, fireballs slow down (and hence isotropize) more quickly for lower jet energies.

\begin{figure*}
\includegraphics[width=9cm]{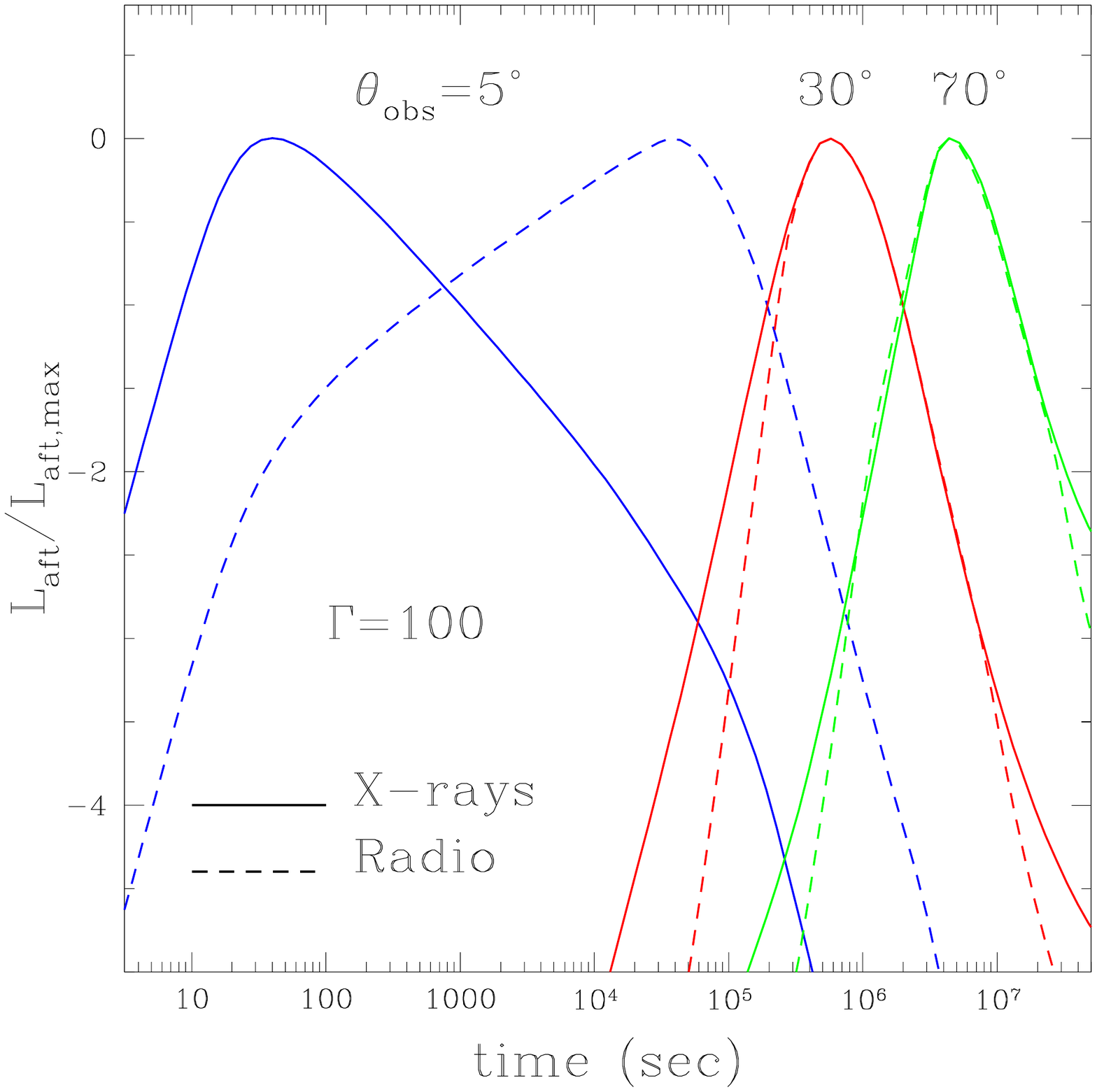}
\includegraphics[width=9cm]{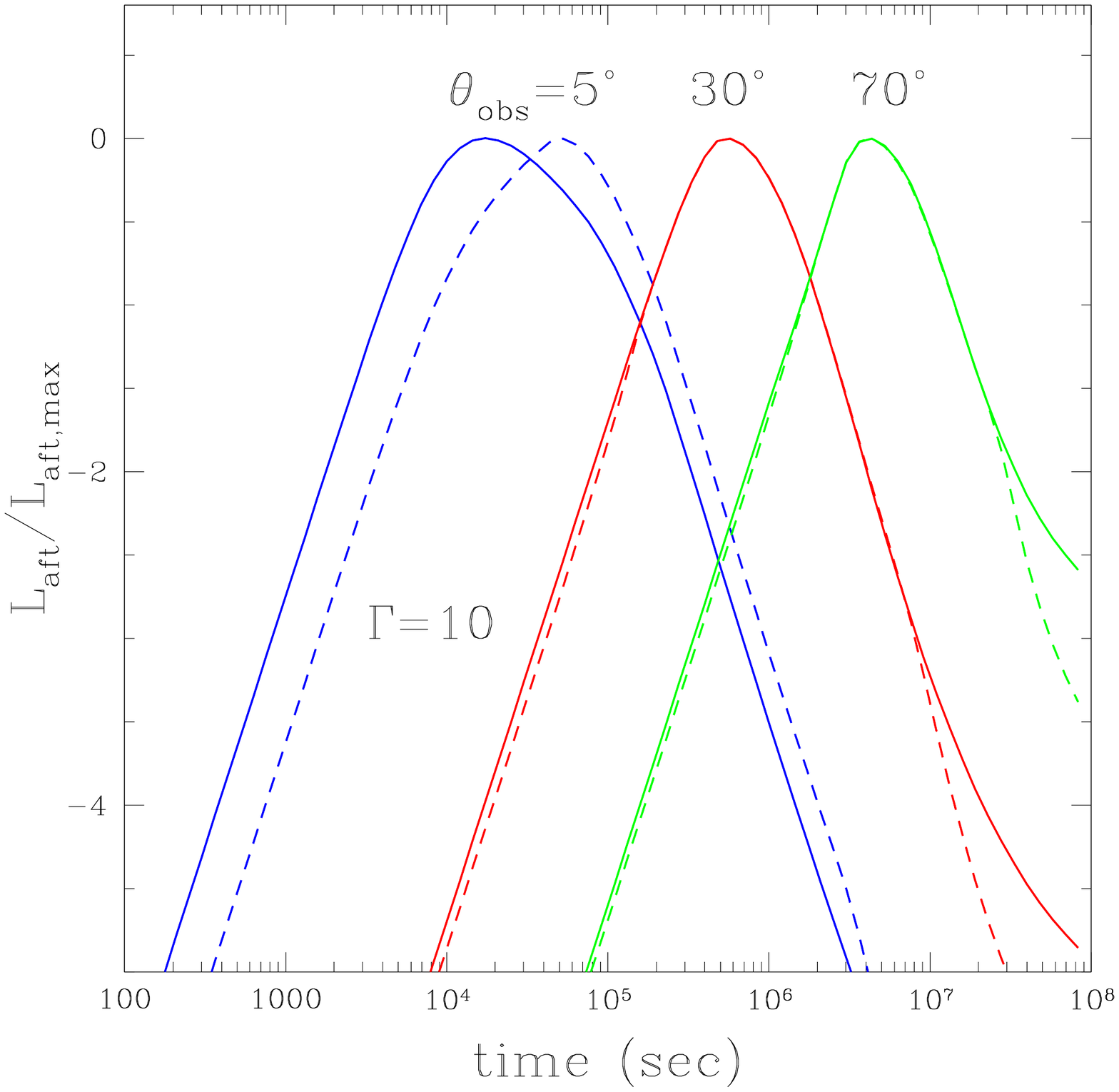}
\vspace{-0.4in}
\caption{Afterglow luminosity as a function of time, for three viewing angles, two energy bands, and the 
high and low $\Gamma$ values under study. In all the cases, the jet energy is $E_{\rm 47}$. When all the other parameters are held fixed, the trend with increasing jet energy
is of light curves peaking at later times. }
\vspace{0.3in}
\label{fig:aft_time}
\end{figure*}

The trend of the afterglow luminosity with time is displayed in Fig.~\ref{fig:aft_time}. For viewing angles larger than $\theta_{\rm jet}$,
the rise of the luminosity with time is essentially determined by the viewing geometry, and hence has very little dependence on frequency: as the fireball slows down and the radiation becomes more isotropic, a larger fraction of the jet emission enters into the line of sight. The maximum is reached when the full jet emission reaches the observer, after which it declines following the further slow down and dimming of the jet.  
On the other hand, for $\theta_{\rm obs}\lesssim \theta_{\rm jet}$, the temporal dependence of the light curve is determined by a combination of factors, and is more strongly dependent on the speed of the fireball, and on the values of the various afterglow parameters ($E,n,\epsilon_e,\epsilon_B$; see \citealt{Sari1998}). As a result, it also has a much stronger dependence on the frequency of the emission, as can be seen in Fig.~\ref{fig:aft_time}.
For all the other parameters fixed, the general trend with increasing energy (not shown here) is of the peak of the light curve to shift to later times. This is due to the fact that the larger the energy, the longer it takes for the fireball to slow down, and hence for the entire jet to come into the line of view of the observer.

\section{Redshift and inclination distribution of the GW-detected BBH mergers}

The redshift distribution of the BHs which are detected via their
mergers in GWs is just beginning to emerge. Since GWs measure the
luminosity distance, they can constrain the redshift dependence of
their sources. However, due to the fact that the detection efficiency
of GW detectors is a function of the BH masses, the redshift
distribution must be fit for simultaneously with that for the BH
masses.

This was done by \citet{Fishbach2018} using the data from the first
six BBH GW events detected by LIGO: GW150914, LVT151012, GW151226,
GW170104, GW170814, and GW170608 \citep{Abbott2016d,Abbott2016a,Abbott2016b,Abbott2016c,Abbott2016e,Abbott2017a,Abbott2017b}.
They used the following parameterization for the distribution of the
primary and secondary masses in the source frame, $m_1$ and $m_2 < m_1$, 
\begin{multline}
{\cal P}\left(m_1,m_2|\alpha,M_{\rm max}\right)\, \\ =\,A\frac{m_1^{-\alpha}}{m_1\,-\,M_{\rm min}}
{\cal H }\, \left(M_{\rm max}\, -  \, {m_1} \right)\, {\cal H} \, \left( m_2\, - \, M_{\rm min} \right),
\label{eq:probmass}
\end{multline}
where $\cal{H}$ is the Heaviside step function, $A$ a normalization factor,
and $M_{\rm min}$ and $M_{\rm max}$ the minimum and maximum BH  mass. Conditioned on the mass of the primary, $m_1$, the mass of the secondary, $m_2$, is drawn from a uniform
distribution between $M_{\rm min}$ and $m_1$.

For simplicity and to avoid over-parametrization, the mass distribution for both
$m_1$ and $m_2$ was assumed to be redshift-independent  (likely a reasonable
assumption given the low-redshift  of the LIGO horizon), so that the full
probability distribution can be written as ${\cal P}(m_1,m_2,z)={\cal P}(m_1,m_2){\cal P}(z)$.  \citet{Fishbach2018} parameterized the redshift distribution as 
\begin{equation}
\label{eq:probz}
    {\cal P}(z) \propto \frac{\mathrm{d} V}{\mathrm{d} z} \left(1 + z\right)^{\gamma - 1},
\end{equation}
where $V$ is the comoving volume \citep{Hogg1999}; 
for a merger rate that is uniform in the comoving frame the parameter $\gamma = 0$ (the distribution of observed redshifts follows $(1+z)^{\gamma - 1}$ due to the redshifting of time between the source and observer frames).  A merger rate that tracks the star formation rate at low redshift corresponds to $\gamma \simeq 3$.

Here we choose the parameter values $\alpha = 1$ and $\gamma = 3$.  We fix the minimum mass in Eq.~(\ref{eq:probmass} ) to $M_{\rm min}=5\,M_\odot$, and the maximum mass $M_{\rm max} = 40 \, M_\odot$.  We fix the $z=0$ BBH merger rate to $100 \, \mathrm{Gpc}^{-3} \, \mathrm{yr}^{-1}$.  To determine which merger events are detected, we use the (semi) analytic selection function described in \cite{Abbott2016b,Abbott2016bs}, with estimated detector sensitivities taken from \citet{Abbott2018LR}.  These choices of parameters are consistent with the inference in \citet{Fishbach2018} and also with the complete set of LIGO observations to date \citep{Abbott2018a}.

For given BH masses and redshifts of the merging binaries, the
inclination angle $\theta_{\rm incl}$ that the perpendicular to the
orbital plane makes with the observer line of sight is computed
according to a probability distribution which is intrinsically
isotropic, but weighed by the LIGO sensitivity to detecting GWs for
various inclinations\footnote{The Monte Carlo simulation drawn from
  the model described in this section can be found here:
  http://www.astro.sunysb.edu/rosalba/LIGOmod/O1+O2unbiased.dat. }.

\section{Event statistics from Monte Carlo simulations}

We use the distributions in \S2 and 3 to generate our event population.

If there is a disk/torus of matter surrounding the merging BBHs, a jet is expected to be launched in the direction perpendicular to its plane (\citealt{Khan2018}; see also \citealt{Yamazaki2016}). The next question is then how the plane of the disk is related to the orbital plane of the merging BHs. The simplest and most natural assumption would be of the two planes to be the same, and hence we perform one set of Monte Carlo simulations considering this scenario, which implies $\theta_{\rm obs}$  to be the same as $\theta_{\rm incl}$. 
However, depending on the source of the matter, this may not necessarily be the case. If, for example, the disk is the remnant of fallback from the SN explosion of one of the two BHs, then its plane would rather be related to the rotation axis of the progenitor star, and hence to the spin of the remnant BH.
Therefore, to account for more general and less restrictive astrophysical scenarios, we additionally perform a second set of simulations in which the viewing angle $\theta_{\rm obs}$ with respect to the jet axis is randomly generated on the sky, and independent of $\theta_{\rm incl}$.  

Given the redshift of the merger event, and the viewing angle with the jet axis simulated according to either of the scenarios above, 
the corresponding electromagnetic luminosity in representative bands
is then computed as described in \S2, and used to calculate the
corresponding fluxes at the observer.

\begin{figure*}
\includegraphics[width=18cm]{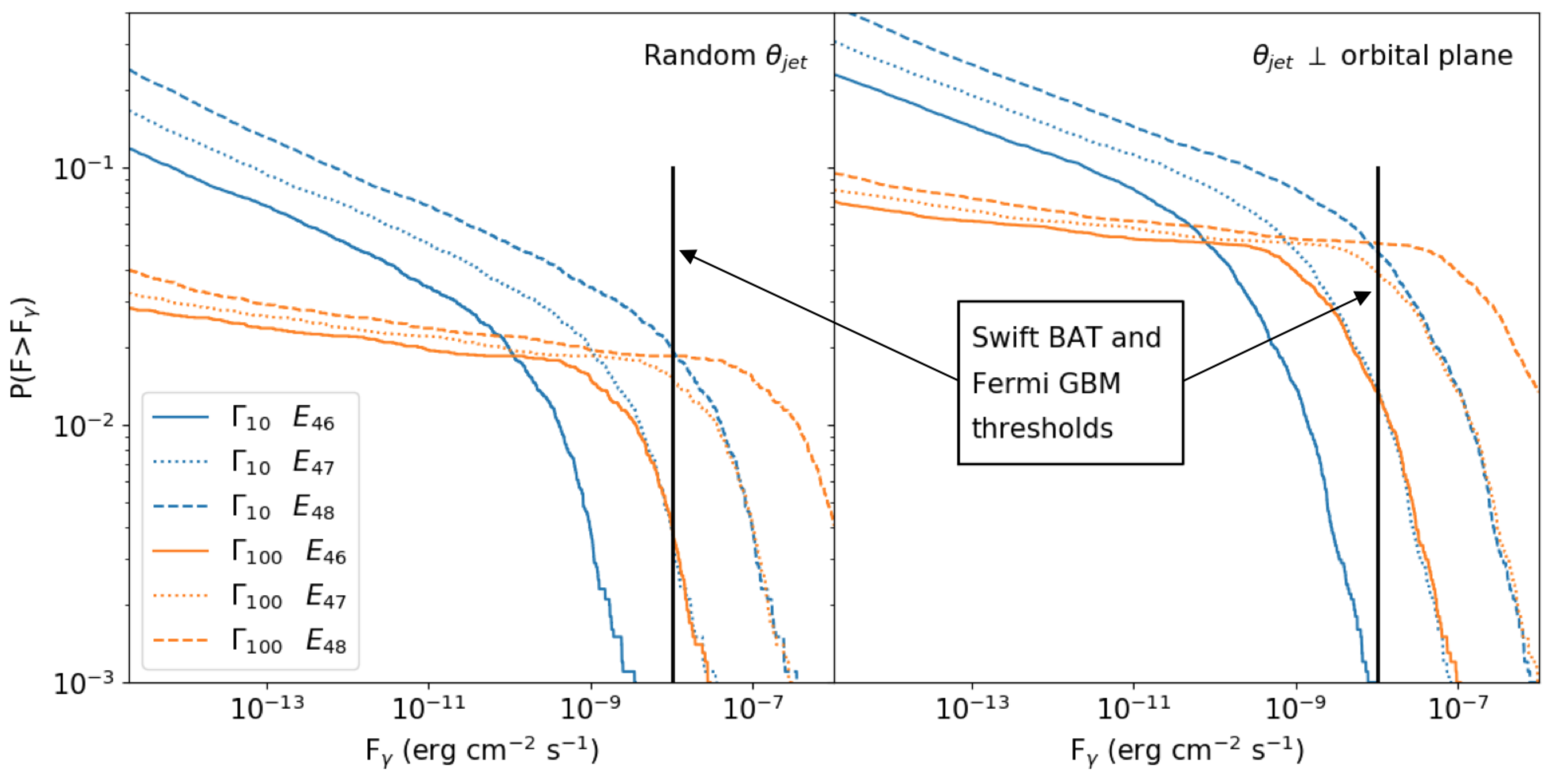}
%\includegraphics[width=9cm]{prob_gamma_mod2.pdf}
%\vspace{-0.4in}
\caption{ Probability of detecting a $\gamma$-ray signal with flux
  larger than $F_{\gamma}$ for each of the models studied here: {\em
    Left:} The jetted EM emission is assumed to be uncorrelated with
  the orbital plane of the BBH prior to merger; {\em Right:} The jet
  axis is assumed to be perpendicular to the orbital plane.  The arrow
  indicates the fraction of events which are above the flux limit of
  Swift/BAT and Fermi/GBM, { before correcting for the limited field
    of view of each instrument.  Note that the BAT and GBM
    sensitivities are comparable but not exactly the same (see text
    for specific details).  }}
\vspace{0.3in}
\label{fig:prob_gamma}
\end{figure*}

For each of the six models discussed in \S3, we ran $N_{\rm
  sim}=10^{4}$ realizations.  The distribution of peak fluxes in
$\gamma$-rays is shown in Fig.~\ref{fig:prob_gamma}, both for the
model with random $\theta_{\rm jet}$ (left panel), as well as for the
one with $\theta_{\rm jet}=\theta_{\rm incl}$ (right panel). The
vertical line marks the flux limit of
$10^{-8}$~erg~s$^{-1}$~cm$^{-2}$.  {This roughly corresponds to the
  sensitivity of Swift/BAT to a typical GRB
  spectrum\footnote{https://swift.gsfc.nasa.gov/about\_
    swift/bat\_desc.html}. In the case of Fermi/GBM, since the
  sensitivity is provided in photon
  counts\footnote{https://gammaray.msfc.nasa.gov/gbm/instrument/} (and
  the conversion to fluence requires a spectral assumption), we simply
  note that the least fluent GRB detected by Fermi has a flux of
  $2.2\times 10^{-8}$~erg~cm$^{-2}$~s$^{-1}$ \citep{Bhat2016}. This is
  a bit higher than the Swift threshold, but of the same order of
  magnitude, which is why for simplicity we indicated the two
  thresholds with the same line in the figure.  Additionally, note
  that the event detection probabilities at those instrumental
  sensitivities need to be corrected for the field of view of each
  instrument (1.4~sr for { Swift} and 9.5~sr for { Fermi).}}

As expected, the probability is somewhat higher in the case in which the jet producing the EM emission is aligned with the orbital angular momentum of the merging BHs. This is because LIGO has an enhanced detection probability to 'face-on' events, which in this case would correspond to jets viewed on-axis.
Among the models explored here, the probability is larger than $\sim 0.1\%$
except for the $\Gamma_{10}E_{46}$ case and randomly-oriented jet. The maximum probability approaches $\sim 6-7\%$ for the model with the largest energy, $E_{48}$, and jet perpendicular to the orbital plane.
The shape of the probability curves is strongly dependent on the fact that, even in mildly relativistic shocks, the brightness is a strong function of the viewing angle (see Fig.~\ref{fig:prompt}), since the high energy $\gamma$-rays are expected during the very early times, on a timescale of seconds, when the fireball has not started to slow down yet.
Hence, to first order, the main determinant of the probability function for visibility in $\gamma$-rays at low fluxes is the luminosity function of the jet. This is especially so for larger $\Gamma$, when the jet side emission drops 
very rapidly to negligible levels. This is why the probability curves for the $\Gamma=100$ cases do not increase significantly  (and are flatter than those for $\Gamma=10$) at the low flux limits displayed in Fig.~\ref{fig:prob_gamma}. 
The high-flux tail of the probability distribution, on the other hand, is dominated by the bright bursts which are seen face-on. The bright tail hence follows the Euclidean $P(>F_{\gamma})\propto F_{\gamma}^{-3/2}$. 

The situation is more complex at longer wavelengths and longer times, as it can be evinced from Fig.~\ref{fig:aft_teta} and \ref{fig:aft_time}. The visibility increases with time at larger angles as the jet decelerates. However, the afterglow luminosity is the brightest at a time which is determined, to first order, by the viewing geometry, but also depends somewhat on the initial $\Gamma$, jet energy, and band (see Fig.~\ref{fig:aft_time} and discussion in \S3).\footnote{There is also a minor dependence on the density of the ambient medium, as well as on the efficiency parameters $\epsilon_e$ and $\epsilon_B$, which we have fixed here, while focusing on exploring the dependence on the parameters with the strongest effect on the model.}. 

\begin{figure*}
\includegraphics[width=9cm]{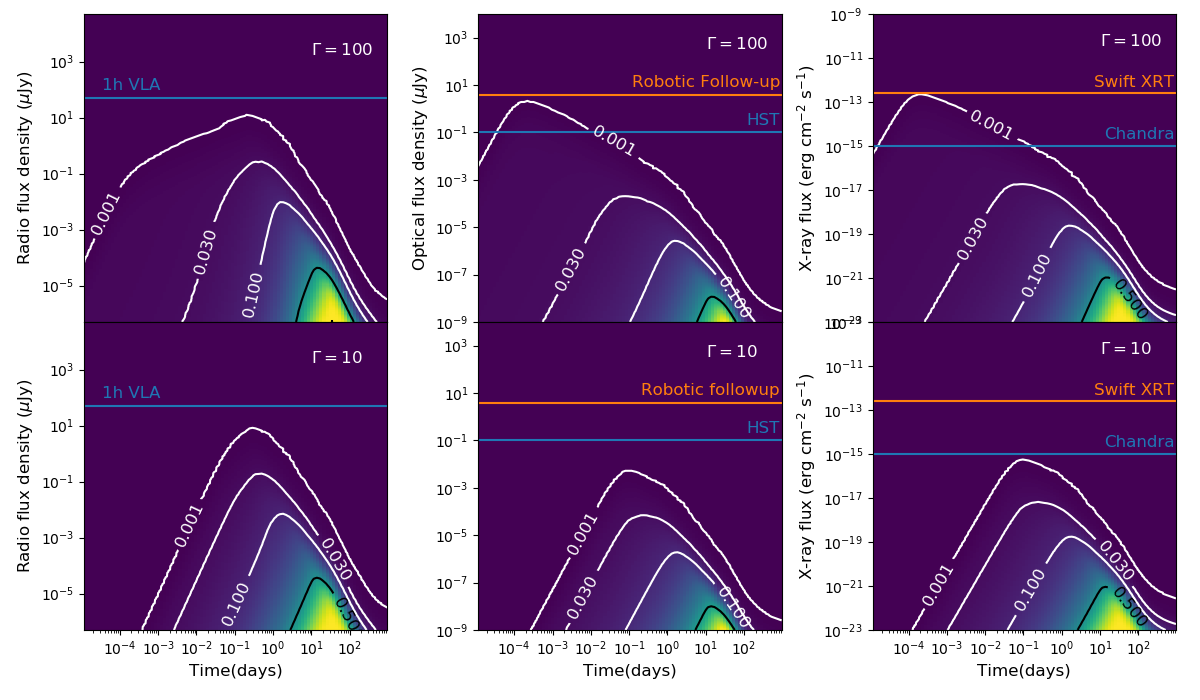}
\includegraphics[width=9cm]{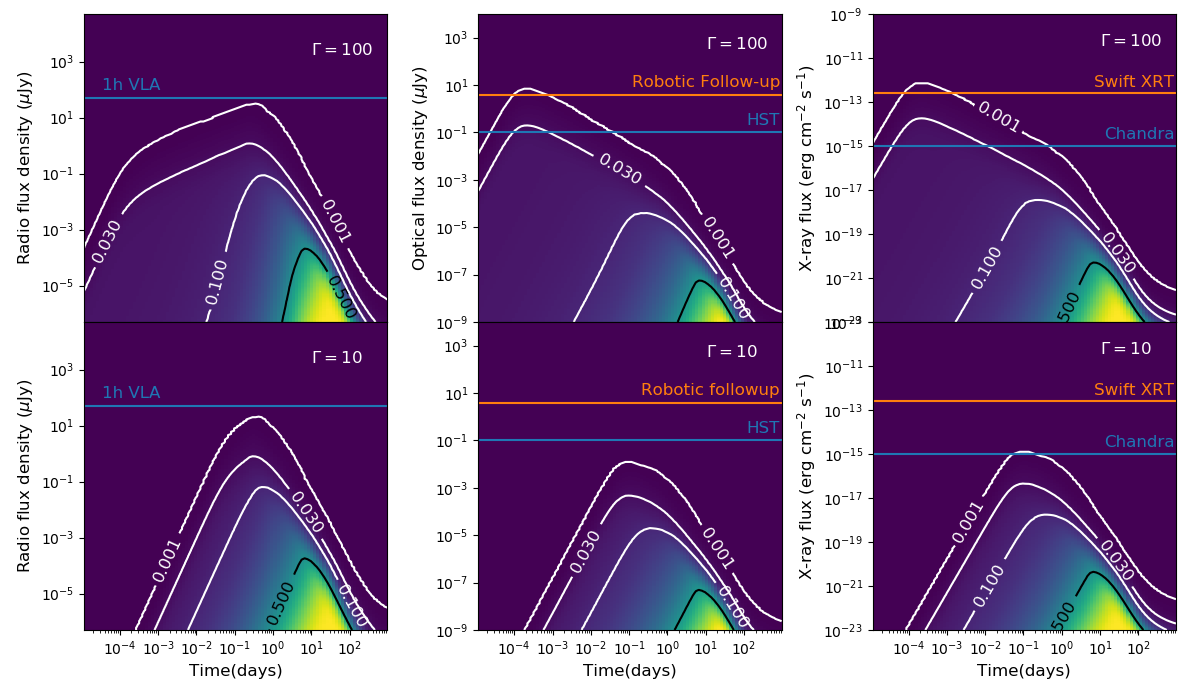}
%\vspace{-0.4in}
\caption{ Probability of detecting an EM counterpart in three representative bands (Radio, Optical, X-rays from left to right of each panel), for the model with jet energy $E_{46}$. {\em Left:} The jet direction is assumed   to be uncorrelated with the orbital plane of the merging BHs. {\em Right:} The direction of the jet is assumed to be the same as the one of the orbital angular momentum. Also included in the figure some representative detection   limits with current observational facilities.
We remind that these probabilities have been computed for a number density of the external medium $n=0.01$~cm$^{-2}$, and that the flux scaling goes from $\propto n^{5/14}$ for a fully radiative blastwave to $\propto n^{1/2}$ for a fully adiabatic one. Hence mergers in dense regions can be expected to have significantly higher detection probabilities.  }
\vspace{0.3in}
\label{fig:prob_AftE46}
\end{figure*}

\begin{figure*}
\includegraphics[width=9cm]{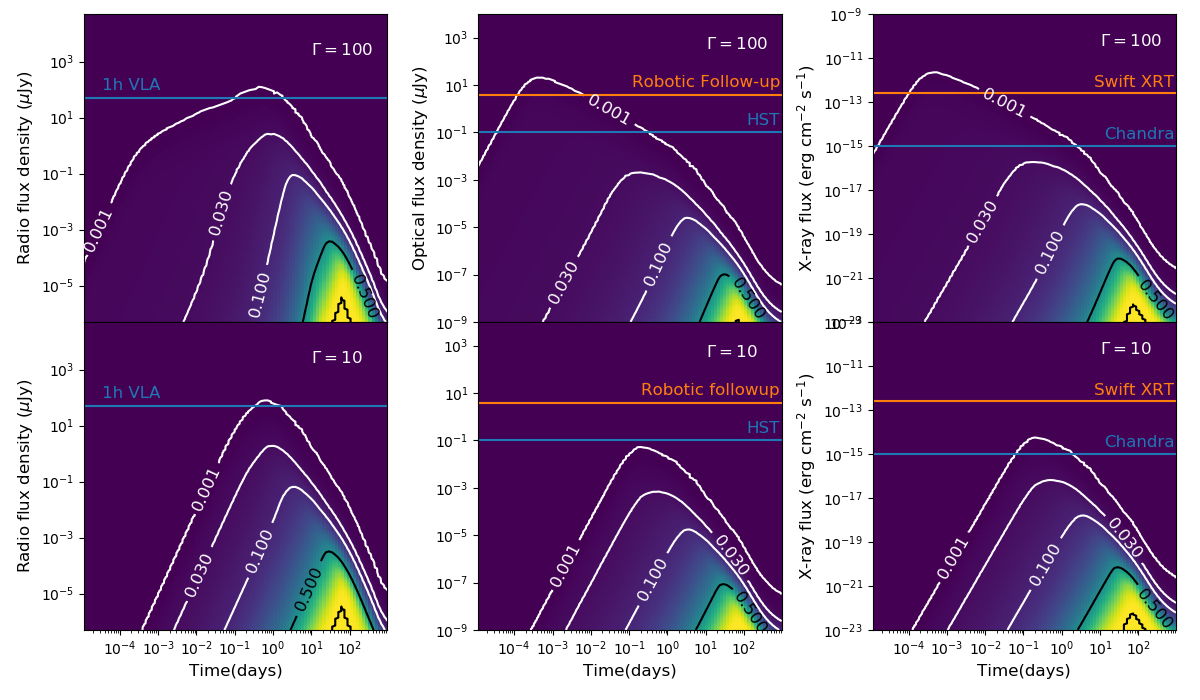}
\includegraphics[width=9cm]{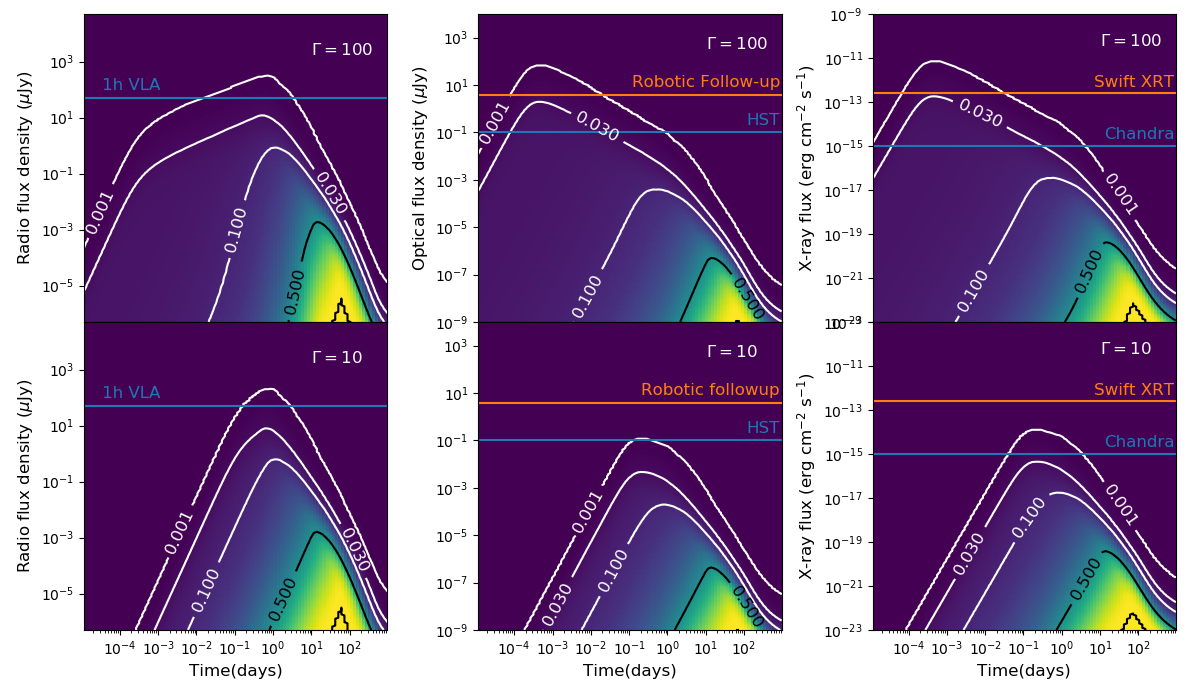}
%\vspace{-0.4in}
\caption{Same as in Fig.\ref{fig:prob_AftE46} but for the model with $E_{47}$. }
\vspace{0.3in}
\label{fig:prob_AftE47}
\end{figure*}

\begin{figure*}
\includegraphics[width=9cm]{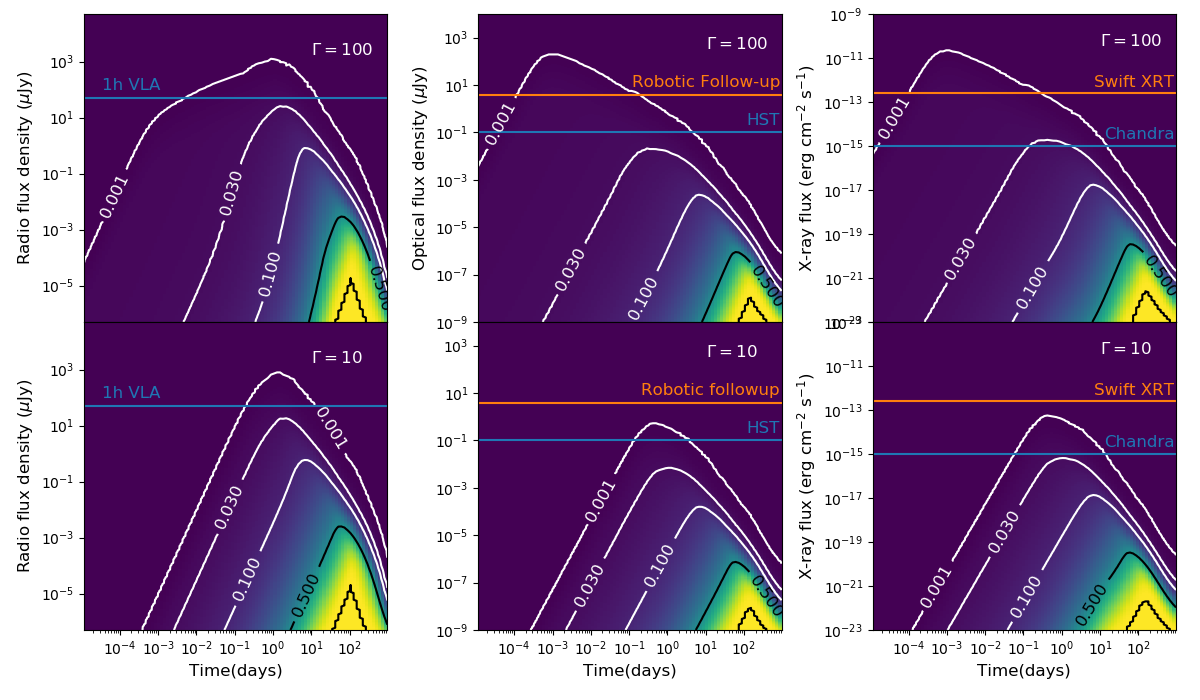}
\includegraphics[width=9cm]{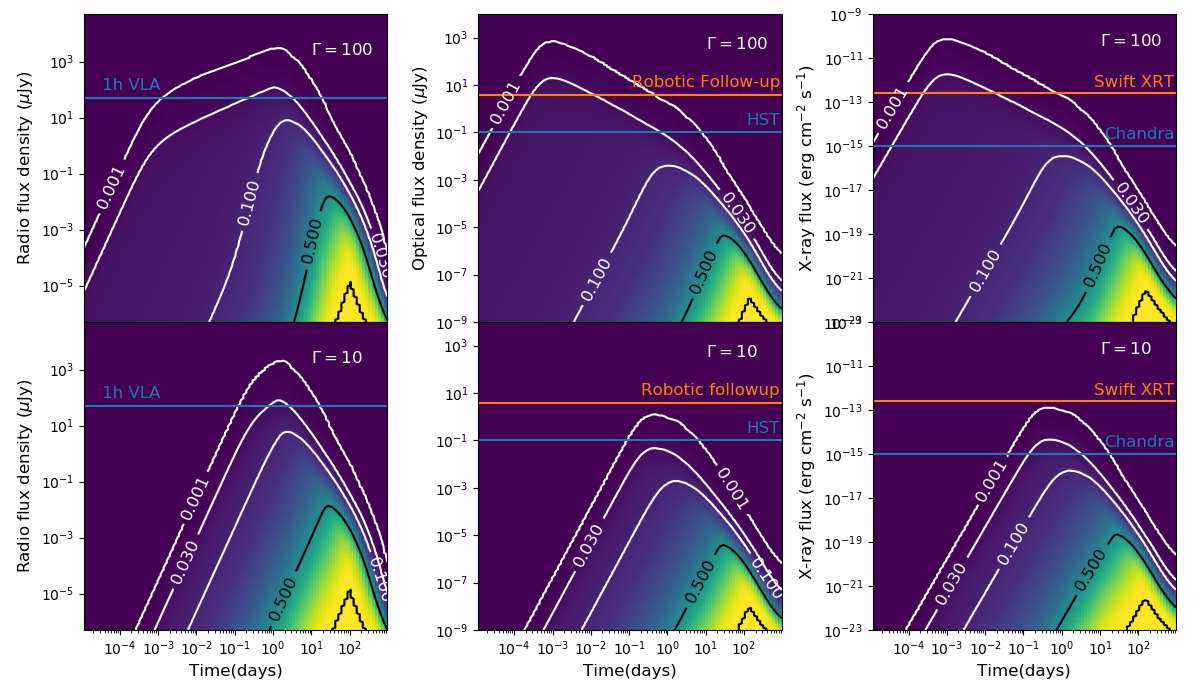}
%\vspace{-0.4in}
\caption{Same as in Fig.\ref{fig:prob_AftE46} but for the model with $E_{48}$.}
\vspace{0.3in}
\label{fig:prob_AftE48}
\end{figure*}

The results of the Monte Carlo simulations for the afterglow in 3 representative bands (X in the 2-10~keV band, Optical at $4.3\times 10^{14}$~Hz and Radio at 1.4\,GHz) are
summarized in Figs.~\ref{fig:prob_AftE46}, \ref{fig:prob_AftE47}, and  \ref{fig:prob_AftE48}.
For each band, we have also included some reference instrumental detection  sensitivities for some of the most common follow-ups in that band.
The detection probabilities are time-dependent, and hence they are influenced by the 
total duration and frequency  of the coverage.
As such, detailed inferences from a comparison between theory and data can only be made case-by-case. Hence in the following we will draw some general conclusions. 
For jet energies $\sim 10^{46}$~ergs (corresponding to $E_{\rm iso}\sim 6.5\times 10^{47}$~ergs), the detection probability with current instruments is practically negligible in any band, even if caught around the time at which the emission peaks in that band, The most optimistic scenario studied here is the $\Gamma_{100}E_{48}$ one, with the jet aligned with the orbital angular momentum (right panel of Fig.~\ref{fig:prob_AftE48}). In all the bands, the detection probabilities turn over at around 3\% for observations carried out around the maximum brightness. This is typically achieved at early times, $t_{\rm obs}\lesssim 1$~day. 

Last, a word of caution in strictly interpreting the probabilities in Figs.~\ref{fig:prob_AftE46},  \ref{fig:prob_AftE47} and \ref{fig:prob_AftE48}.
We remind the reader that the afterglow luminosity depends on several microphysical parameters, as well as on the ambient medium density, as discussed in detail in \S2.  
Therefore, each of the curves showed in Figs.~\ref{fig:prob_AftE46},  \ref{fig:prob_AftE47} and \ref{fig:prob_AftE48} should be interpreted as having a swath of variability for the quoted model parameters. 

\begin{figure*}[!t]
\includegraphics[width=8.5cm]{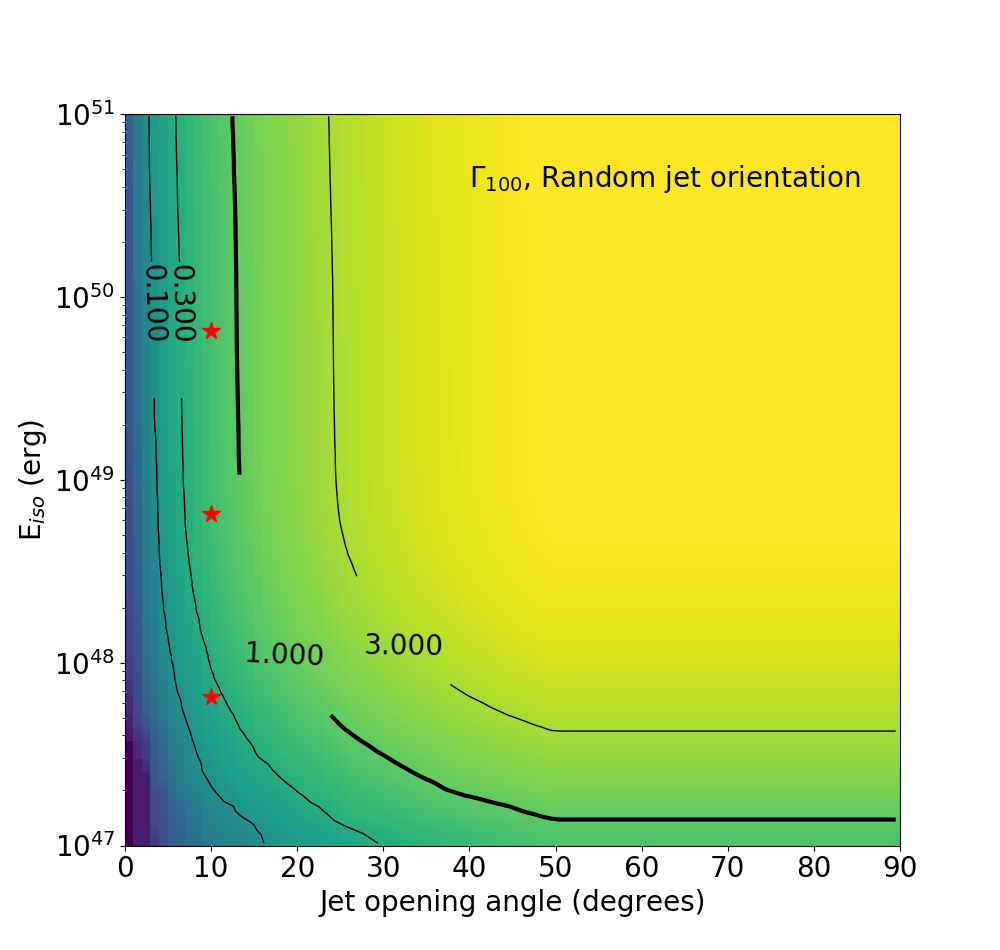}
\includegraphics[width=8.5cm]{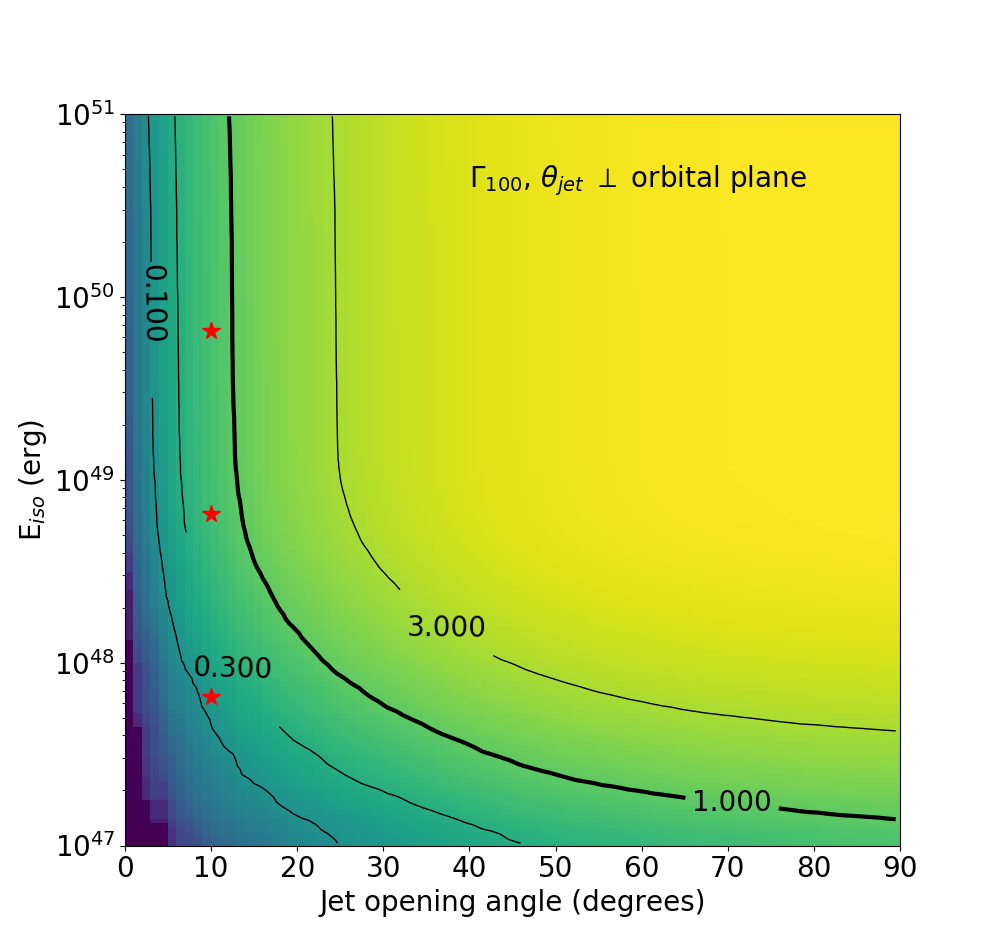}\\
\includegraphics[width=8.5cm]{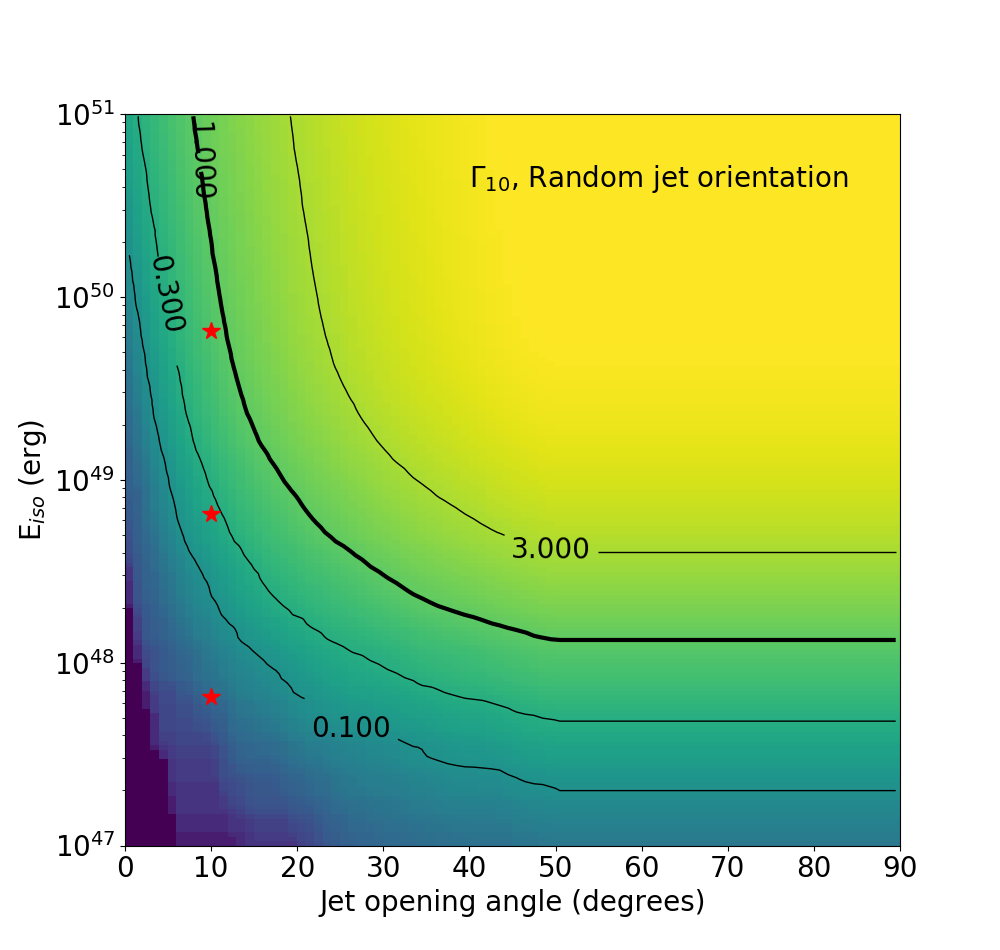}
\includegraphics[width=8.5cm]{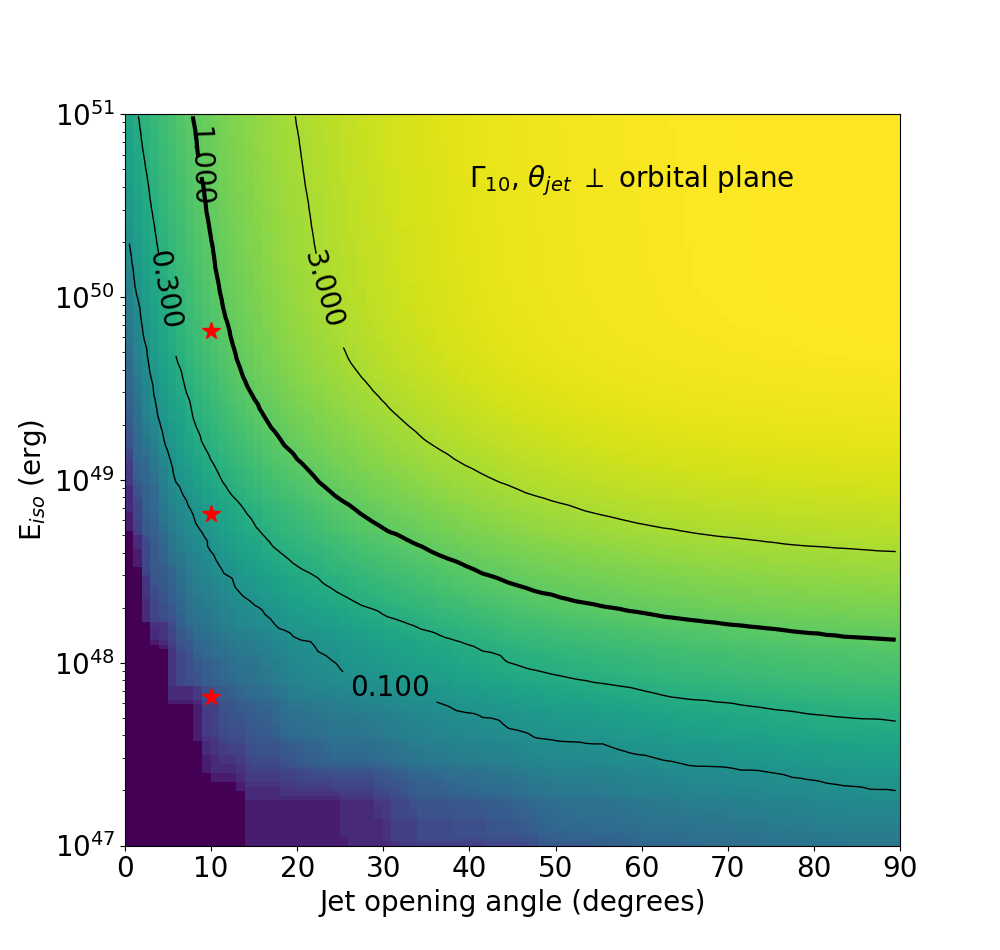}
%\vspace{-0.4in}
\caption{Average number of events with $\gamma$-ray flux above
the Fermi/GBM threshold, out of 10 GW-detected BBH mergers.
The stars indicate the models studied in more detail here.
By allowing at most one $\gamma$-ray detection in the sample,
the permitted parameter space is restricted to the range 
$[(E_{\rm iso}/10^{48}{\rm erg})(\theta_{\rm jet}/20^\circ) -
(E_{\rm iso}/10^{49}{\rm erg})(\theta_{\rm jet}/20^\circ)]
\lesssim 1$, with the specific value
dependent on the Lorentz factor and
on the relative inclination of the jet with respect to the orbital plane of the merging BBHs.}
%\vspace{-0.8in}
\label{fig:prob_Eteta}
\end{figure*}

\section{Constraints on emission models from EM follow-ups to LIGO BBHs mergers to date}

LIGO and Virgo have detected 10 BH-BH mergers during the first two observing runs O1 and O2 \citep{Abbott2019a}. Among all the models studied here, the probability of detecting  EM emission in $\gamma$-rays above the Fermi/GBM threshold is the largest ($\sim 5-7$\%) for $\theta_{\rm jet}\perp$ orbital plane, and
the highest energy case
($E_{48}$), as well as 
for $E_{47}\Gamma_{100}$. Thus, for 10 events with energetics and $\Gamma$ factors in that range, we would expect to see an average of $\sim 0.5$ event.  Therefore the tentative detection of one counterpart in $\gamma$-rays would not be surprising. As a reference, recall that the inferred isotropic energy of that event is $10^{49}$~ergs, which translates into a jet energy of $10^{49} (1-\cos\theta_{ \rm jet})$. For $\theta_{\rm jet}=10^\circ$, this is $1.5\times 10^{47}$~erg. 

We further generalize the above	constraint by running an extended series of
Monte Carlo simulations for a wider range of jet angles and isotropic
equivalent energies; for each combination, we compute the average
number of events (out of 10 GW-detected BBH mergers) with $\gamma$-ray
flux above the Fermi/GBM detection threshold.  The results are
reported in Fig.\ref{fig:prob_Eteta}, for both the low and high
$\Gamma$ models, as well as the two scenarios for the direction of
$\theta_{\rm jet}$. The	stars indicate the three $[E_{\rm
iso}\theta_{\rm jet}]$ combinations studied in more detail here.  As
expected, larger energetics require smaller jet angles in order not to
overpredict the number of events with a $\gamma$-ray detection.
Allowing the number of detections to be no more than 1 out of 10, we can
already restrict the permitted parameter space to the range
$[(E_{\rm iso}/10^{48}{\rm erg})(\theta_{\rm jet}/20^\circ) -
(E_{\rm iso}/10^{49}{\rm erg})(\theta_{\rm jet}/20^\circ)]
\lesssim 1$, with the specific value
dependent on the Lorentz factor and
on the relative inclination of the jet with respect to the orbital plane of the merging BBHs.

At longer wavelengths, there have been no reported EM counterparts from follow ups to the first 6 BBH merger events.
Assuming that the situation remains the same after the complete follow up catalogue has been published, the probability of one detection in the best case scenario and with continuous follow up is  at most
$\sim 30\%$.  
Therefore, the current (afterglow) data set cannot be used as yet to rule out jet production within the range of models studied here.
However, we remind once again that our afterglow models have been run for a fixed density value ($n=0.01$~cm$^{-3}$).
Merger events in denser regions would have brighter emission than the one computed here, and hence  detection probabilities have a degree of degeneracy between the source properties and the ambient ones.  
To allow the community to use our results to restrict the allowed parameter space by means of each new followup in some energy band and at some specific observing time, we have put our current models online\footnote{http://www.astro.sunysb.edu/rosalba/EMmod/models.html}, and we are populating them further with models run with a wider range of parameters.

\section{Summary and Discussion}

The detection of GWs from binary black hole mergers has provided yet
another confirmation of the theory of General Relativity.
However, the tentative detection of an EM counterpart to GW150917
\citep{Connaughton2016}  had not been predicted by any theory, and
hence it gave rise to a number of ideas of different nature, from the
mundane to the exotic. If EM radiation from merging BHs was detected
at high confidence, it would thus revolutionize our pre-concepts of
merging BHs.

While the energy production mechanisms proposed for EM emission
are diverse, a common feature to a sudden release of energy is
the formation of a relativistic shock which plows into the
medium, giving rise to radiation spanning a wide electromagnetic
range, from $\gamma$-rays to radio. Both simulations and
observations of this phenomenon have shown that the radiating
outflow is jetted. Emission at wide angles is dominated
by interaction of the jet with surrounding dense material, as
demonstrated by the binary NS merger GW170917; in this case
the interacting material is provided by the tidally disrupted
matter of the neutron stars. Such ejecta is however  not present
in the case of a BBH merger, resulting in much weaker
angular emission.

Assessing the detection probability of EM emission is of paramount
importance in order to be able to extract meaningful information as
more data are gathered from EM followups to BBH mergers.  To this aim,
here we have performed a Monte Carlo simulation of a population of BBH
mergers, with a redshift distribution derived from the current
observed population, and for a range of energies and Lorentz factors
of possible jets driven at the time of the merger. The angular
emission of the jet in different energy bands has been numerically
calculated for each event as a function of time.

Among the models which we explored in detail, we find that, in $\gamma$-rays, the detection probability with the
Swift/XRT and Fermi/GBM bands is up to  $\sim $6-7\%  in the model with the largest jet energy, $E_{48}$, and with the
direction of the jet aligned with the orbital angular momentum 
of the merging BHs (since
these events are more easily detectable by LIGO). The probability for detection in  $\gamma$  is largely determined
by the angular size of the jet, since at early times the
high Doppler factor largely suppresses the side emission. Hence,
to generalize and further explore the consequences of our results for $\gamma$-ray followups to date, we additionally ran a grid of models for a much larger range of jet energies and jet opening angles. The condition that $\lesssim 1$ event out of 10 GW-detected BBH mergers is above the Fermi/GBM threshold imposes that any currently allowed emission model has to satisfy the condition 
$(E_{\rm iso}/10^{49}{\rm erg})(\theta_{\rm jet}/20^\circ)\lesssim 1$ for the most favorable scenario.

At longer wavelengths, the detection probability in each band becomes
time-dependent, and the precise time of the maximum depends on a combination of the
fireball parameters, such as its energy and initial Lorentz
factor, and the viewing angle to the observer.
However, even in the best case model studied here,  
the detection probability with current observational facilities in typical observation bands (X, O, R) is at most around 3\%. 
Early followups, within a day, yield higher chances of catching the afterglow radiation.

Lack of detection of X-ray through radio emission  from the first 10 LIGO/Virgo events (assuming that the complete catalogue has the same properties of the first 6 events) remains still unconstraining for the range of models studied here, since the detection probability would be at most $\sim 30\%$. More events are needed before being able to put more  stringent constraints down to the energy levels considered here (unless jets were considerably wider than $\sim 10^\circ$ and/or the ambient density of the medium in which the merger occurred was on average much larger than  $\sim 0.01$~cm$^{-3}$). 

Finally, note that the EM detection probabilities that we have calculated here have assumed the detection sensitivity of LIGO/Virgo in runs O1/O2. As the sensitivity to GW detection improves in future runs, the probability of observing EM radiation in followups to GW-detected BBH mergers {\em decreases}, since more events will be detected from higher redshifts, and hence they will appear on average dimmer to the observer (for the same model parameters). The total number of GW events (and hence potential follow ups) will however increase, and hence it will be important to have a wide range of models to compare against. While here we have reported and discussed the results of a few representative cases, we are building a public, online library of models in several representative bands for a much wider range of parameters. Our library (see footnote 6 for location) will allow to use each new limit on EM emission to carve out a region of highly disfavored emission models, so that, over the years to come, we will learn how dark BBHs mergers actually are. Future missions with higher sensitivity than the current facilities, such as the James Webb Space Telescope, will play a crucial role in this pursuit.

\acknowledgments
RP acknowledges support from the NSF under grant AST-1616157.
The Center for Computational Astrophysics at the Flatiron Institute is supported by the Simons Foundation.

\bibliographystyle{aasjournal}
\bibliography{biblio}

\end{document}